\documentstyle[12pt,psfig]{article}
\begin{document}

\title {\textsf{Phenomenon of Gamma-Ray Bursts \\ as Relativistic Detonation of
Scalar Fields}}
\author { V.~Folomeev,
V.~Gurovich \thanks{email: astra@freenet.kg} \enspace and
R.~Usupov
\\
{\small \it  Physics Institute of NAN KR,}\\
         {\small \it 265 a, Chui str., Bishkek, 720071,  Kyrgyz Republic.}}
\date{\small \it (21 August 2000)}
\maketitle

\begin{abstract}
\noindent In the modern Universe the existence of various forms of
scalar fields is supposed. On the one hand these fields can
explain recently discovered positive $\Lambda$-term(see e.g.
Ref.~\cite{ref:Sahni}), on the other hand its form cluster systems
creating gravitational wells for galaxies and their clusters. At
that a natural hypothesis is the existence of compact
configurations ("stars") from scalar fields with a large enough
energy density and total mass. The hypothesis is that the energy
of these fields can be converted in relativistic plasma by an
explosive way. Such process can be initiated by collision of
relativistic particles which form a relativistic microscopic
fireball. Thus effective temperature can amount to value
sufficient for change of phase for scalar fields. Then the wave of
relativistic "detonation" similar to the same process in classical
physics will be spread from this source. In this paper the
parameters of such field star and process of detonation are
estimated. If the effect of the indicated change of phase (or
something similar to one) exists, it is possible to get the
parameters of relativistic plasma (macroscopic fireball) which
could generate gamma - bursts. If in the modern Universe there is
such unique form of a matter as fields of high density it would be
strange for Nature not to take advantage of the possibility to
convert their energy to radiation by an explosive way.
\end{abstract}

\section{Introduction}

\noindent Now there is no conventional mechanism of origin of
gamma-ray bursts (GRB). Also the accumulation of all required
energy in initial small volume is rather doubtful. At the same
time there are theoretical models in the theory of the early
Universe when the beginnings of large volume of relativistic
plasma occurs fast enough with expansion of front of a spherical
wave (see e.g. an expansion of the bubble at decay of false vacuum
(creation of the Universe from the bubble~\cite{ref:Col})). We are
based on the following assumptions: now the candidate for a dark
matter in the Universe are the scalar fields of a various type.
The mass of these fields can essentially exceed a mass of luminous
matter and then they first form the large-scale structure of the
Universe. An apparent matter concentrates in their gravitational
potential wells. It is possible that these fields can create and
more dense ''star-like'' configurations. Such relativistic models
also were studied for a number of
years~\cite{ref:Ruf,ref:Pir,ref:Miel}. Below will be shown that
the similar star-like configurations with sufficiently large size
and with big mass density are exist at weak gravitation when the
configuration can be described in Newtonian approximation.

Let's imagine that as a result of collision of relativistic
particles some critical size of a fireball is formed (note that
last one is present in other models of GRB also~\cite{ref:Post}).
It is that seed mechanism which effectively transfers scalar
fields in creating pairs of unstable elementary particles. Such
mechanism can be imagined as a beginning of fast oscillation of a
field on exterior boundary of a fireball. Such process is
equivalent to a relativistic detonation. The role of a chemical
energy turning into relativistic plasma at the front of a
detonation wave will be played by the energy of a scalar field
which intensively passes in pairs of creating particles -
antiparticles. Now such fireball, expanding with a relativistic
velocity, will not depend on weak gravitational fields of
Newtonian configuration and it can be considered in special
relativity.

\section{Relativistic Detonation}

\noindent Let's show the self-similar solution for a spherical
relativistic detonation which transfer in the well known solution
of Ya. B. Zeldovich in case of small velocities~\cite{ref:Land}.
The set of equations of relativistic hydrodynamics is convenient
to present in a spherical frame with use of usual
three-dimensional radial velocity of plasma $v$
~\cite{ref:Land,ref:Baum}. Thus the equation of motion looks like:
\begin{equation}
\label{eq1}
\frac{1}{\theta ^{2}}\left( \frac{\partial v}{\partial \tau }+v\frac{%
\partial v}{\partial r}\right) +\frac{1}{w}\left( \frac{\partial p}{\partial
r}+v\frac{\partial p}{\partial \tau }\right) =0
\end{equation}
and the law of conservation of energy:
\begin{equation}
\label{eq2}
\frac{1}{w}\left[ \frac{\partial \varepsilon }{\partial \tau }+v\frac{%
\partial \varepsilon }{\partial r}\right] +\frac{1}{\theta ^{2}}\left( \frac{%
\partial v}{\partial r}+v\frac{\partial v}{\partial \tau }\right) +\frac{2v}{%
r}=0,
\end{equation}
here $\theta ^{2}=1-v^{2}$, $w=\varepsilon +p$\ and $c=1$. As well
as in a nonrelativistic case the motion of plasma behind the front
of the detonation wave is considered as isentropic and the
relevant solution is described only by two equations referred
above. The pairs of relativistic particles which are creating in
area behind the front of wave generate high-temperature plasma
with the equation of state:
\begin{equation}
\label{eq3}
p=\omega ^{2}\varepsilon ;\,\,\,\,\,\,\,\,\,\omega ^{2}=\left( \frac{%
\partial p}{\partial \varepsilon }\right) _{S}=1/3,
\end{equation}
where $\omega $\ is the sound velocity. Similarly to the problem
about a spherical detonation we shall search for a solution
depending on the self-similar variable
\begin{equation}
\label{eq4}
 \xi =r/\tau .
\end{equation}
Thus the set of Eqns. (\ref{eq1}), (\ref{eq2}) passes in the
system of ordinary differential equations and supposes the
obtaining of one equation on $v$:
\begin{equation}
\label{eq5}
\frac{dv}{d\xi }\left[ \frac{1}{\omega ^{2}}\left( \frac{v-\xi }{1-v\xi }%
\right) ^{2}-1\right] =\frac{2v}{\xi }\frac{\theta ^{2}}{1-v\xi }.
\end{equation}
In a nonrelativistic case ($v(\xi ),\,\xi \ll 1$ ) the last
equation passes in classical one~\cite{ref:Land}. The qualitative
analysis of this equation is
similar to the known mentioned result. The solutions for $v$\ and $%
\varepsilon $\ have the infinite derivative at the front of wave (
$\xi =\mathcal{D}$\ is the velocity of the detonation wave). It
follows from the known fact of the detonation theory: the velocity
of plasma, outgoing from the wave front, is equal to the sound
velocity $\omega $. The expression in the parenthesis in the {\it
l.h.s.} of Eqn. (\ref{eq5}) corresponds to the relativistic law of
a velocity addition and at $\xi =\mathcal{D}$ it is equal to
$\omega ^{2}$. The last one means that at tending of argument $\xi
$\ to the wave front the whole expression in brackets aspires to
zero from above. The {\it r.h.s.} of Eqn. (\ref{eq5}) remains
finite that means tending of the derivative to infinity. The
examination of a transformation of the velocity of plasma in a
zero for Eqn. (\ref{eq5}) is completely similar to the analysis of
Ref.~\cite{ref:Land} as the motion becomes nonrelativistic. The
single unknown parameter in this problem is the velocity of the
detonation wave $\mathcal{D}$. In case of classical detonation it
is determined by an internal energy of explosive. For the
considered hypothesis it will be defined by the energy flux
density and by momenta of a scalar field entering the wave front.
\begin{figure}
\begin{picture}(500,320)
\put(-50,-490){\includegraphics{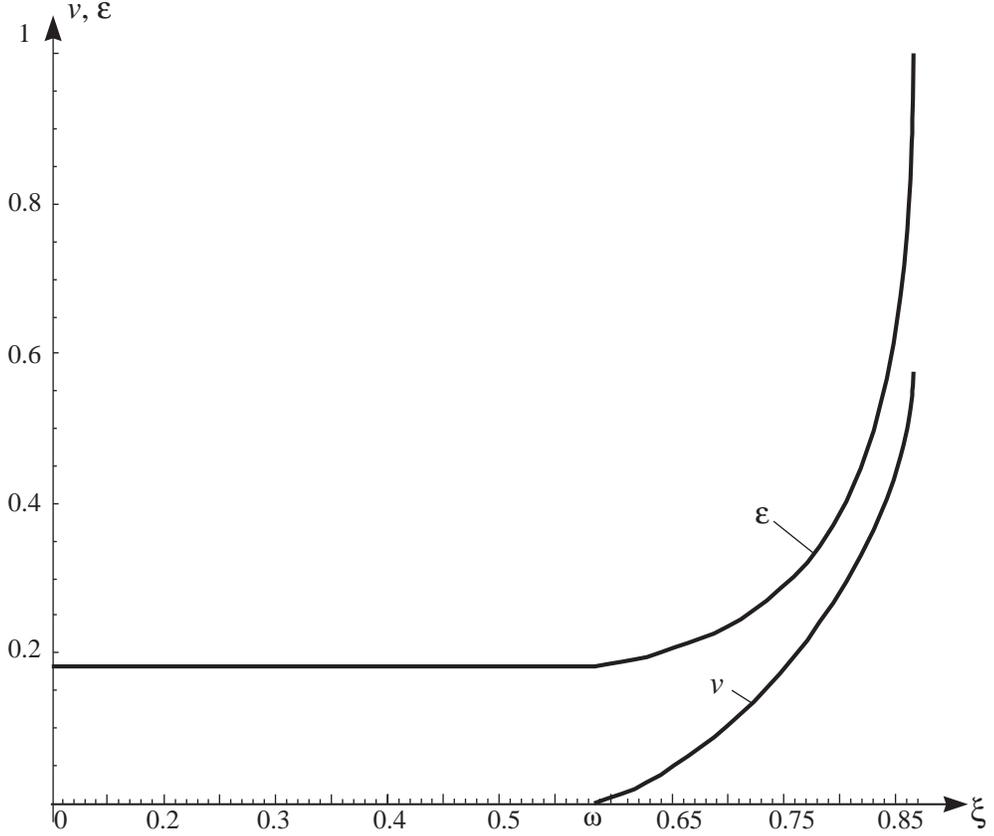}}
\end{picture}
\caption{\it The dependence of the energy density $\varepsilon$
and velocity $v$ of plasma on the self-similar variable $\xi$
behind the front of detonation wave.}
\end{figure}

The discussion of the possible mechanism of ''recycling'' of the
field behind the wave front in relativistic plasma is considered
below. Let's specify here the following estimation of value of
$\mathcal{D}$\ and energy density behind the detonation wave.
Let's consider that the scalar field is set in the simplest
variant:
\begin{equation}
\label{eq6}
T_{i}^{k}=\varphi _{,i}\varphi ^{,k}-\delta _{i}^{k}\left( \frac{1}{2}%
\varphi _{,\mu }\varphi ^{,\mu }-V(\varphi )\right)
,\,\,\,\,\,\,\,\,\,V(\varphi )=m^{2}\varphi ^{2}/2
\end{equation}
and is in scalaron regime (the fast oscillations with frequency $m$): $%
\varphi (r,t)=a(r)\sin mt$. The spatial changing of the field is considered
small. Thus the energy density of such field (in a laboratory frame) is
determined by:
\begin{equation}
\label{eq7}
\varepsilon _{f}=m^{2}a^{2}/2.
\end{equation}
The expression for $\mathcal{D}$\ and energy density of plasma
behind the wave front is
determined from conservation laws $T_{0}^{1}(field)=T_{0}^{1}(plasma)$ and $%
T_{1}^{1}(field)=T_{1}^{1}(plasma)$ for the observer which is in
the rest at the wave front. (Remind that the plasma goes away the
wave front with the velocity $\omega $.) Hence
\begin{equation}
\label{eq8}
{\mathcal{D}}=\frac{2\omega }{1+\omega ^{2}};\,\,\,\,\,\varepsilon _{p}=\frac{2}{%
1-\omega ^{2}}\varepsilon _{f}.
\end{equation}
In case of relativistic plasma $\omega =1/\sqrt{3}$ and ${\mathcal{D}}=\sqrt{3}%
/2,\,\,\,\,\varepsilon _{p}=3\,\,\varepsilon _{f}\,$. The
self-similar solutions for this case are presented in Fig. 1.

\section{About possible mechanism of transition field - plasma}

\noindent The most complicated are the processes at the front of
detonation wave - transition of the field in plasma. Let's note
that for modern space fields (cluster component of the field in
the terminology of Ref.~\cite{ref:Sahni}) the most different
masses of quantums of a scalar field are choose. Let's specify
also that explicitly enough the relativistic models of boson stars
are studied in different variants~\cite{ref:Miel}. The most
beautiful mechanism of the transition field - plasma can be the
change of phase at penetration of the field in temperature bath
with temperature $T$\ close to critical at which there is a
catastrophic decreasing of a mass of quantums of scalar
fields~\cite{ref:Linde}. The strong increasing of the oscillation
amplitude behind the wave front realizes in multiple creation of
pairs of elementary particles and filling of the next layer by hot
plasma. (Other mechanisms of such transition are not excluded
also.)

The most simple explanation of duration of radiation of the GRB
would be the existence of extended enough area filled by scalar
field with energy density and size about 1-10 light second
acceptable to the given mechanism that corresponds to duration of
complete radiation of the GRB. In considered variant it is close
to time of ''burning'' of the field by the detonation wave. As the
known examples of the field stars correspond to relativistic
configurations we think that it is interested to specify a
possibility of existence of scalaron Newtonian stars which
together with high energy (mass) density would have a large enough
expansion commensurable with the mentioned above typical size.

\section{Newtonian Scalaron Star}

\noindent The Newtonian gravitational configuration filled by
scalar field with Lagrangian density is considered
\begin{equation}
\label{eq9}
 L=\frac{1}{2}\varphi _{,i}\varphi ^{,i}-m^{2}\varphi
^{2}/2
\end{equation}
and equation for the field is
\begin{equation}
\label{eq10}
\frac{1}{\sqrt{-g}}\frac{\partial }{\partial x^{i}}\left[ \sqrt{-g}g^{ii}%
\frac{\partial \varphi }{\partial x^{i}}\right] =-m^{2}\varphi ,
\end{equation}
where $m$\ is the mass of the field quantum. In Newtonian
approximation
\begin{equation}
\label{eq11}
 g_{00}=1+2\Phi
/c^{2};\,\,\,\,\,g_{11}=-1;\,\,\,\,\,g_{22}=-r^{2};\,\,\,\,\,g_{33}=-r^{2}%
\sin ^{2}\theta .
\end{equation}
Consider further the stationary configuration in which the field
$\varphi (r,t)=a(r)\sin mt$\ and the gravitational potential $\Phi
/c^{2}=-\Psi (r),\,\,\Psi >0$\ depends only on $r$. Thus the field
equation will has the form:
\begin{equation}
\label{eq12}
\frac{1}{r}\frac{d^{2}}{dr^{2}}(ra)+2m^{2}\Psi a=0.
\end{equation}
The amplitude of the field also has spatial oscillations but by virtue of
smallness of $\Psi $\ the spatial derivatives are much less temporal one.
For this reason the mass density of the field $\varphi $\ is determined by
the formula:
\begin{equation}
\label{eq13} \rho =\varepsilon /c^{2}=\left[ \dot{\varphi
}/2+m^{2}\varphi ^{2}/2\right] /c^{2}=m^{2}a^{2}/2c^{2}.
\end{equation}
It gives the basic contribution to gravitation. The spatial derivatives on $%
a(r)$\ create an effective pressure counterweighting the forces of
gravitation. The potential $\Psi (r)$\ has found from the Poisson equation:
\begin{equation}
\label{eq14}
\Delta \Psi =-\frac{4\pi G\rho }{c^{2}}=-\frac{2\pi
G}{c^{4}}m^{2}a^{2}.
\end{equation}
The set of Eqns. (\ref{eq12}), (\ref{eq14}) determines the
structure of scalaron ''star''. The solutions of the system are
convenient for investigation in dimensionless form. Thus the
typical quantities used at undimensionality give the possible
estimations of values for parameters of such ''star'':
\begin{equation}
\label{eq15} r=\ell\xi ;\,\,\,\,\,\Psi =\Psi
_{0}U;\,\,\,\,\,a=a_{0}x,
\end{equation}

\[
\ell^{2}\sim 10^{9}\frac{n}{m_{eV}\sqrt{\rho _{0}}}\,\,cm^{2};\,\,\,\,\,%
\Psi _{0}\sim 10^{-18}n\frac{\sqrt{\rho _{0}}}{m_{eV}};\,\,\,\,\,a_{0}%
\sim 10^{6}\frac{\sqrt{2\rho _{0}}}{m_{eV}}\,\,\left( \frac{erg}{%
cm^{3}}\right) ^{1/2},
\]
here $\rho _{0}$\ is the central field density in g/cm$^{3}$,
$m_{eV}$\ is the mass of a field quantum in eV, $n \gg 1$\ is the
dimensionless number.

We take e.g. $m_{eV}=0.5\times 10^{-8}$eV\ and central mass
density $\rho _{0}=10^{16}$g/cm$^{3}$. Then at $n=5$\ the typical
size of the configuration will be $\ell \sim 10^{5}$cm and total
mass in the order of $M=10^{31}$g. Thus the potential $\Psi
_{0}\sim 0.1$\ that is evidence of applicability of Newtonian
gravitation theory in this case. If the mechanism of transition of
the field in plasma is realized in the front region then for this
rough estimate we have the following:

According to the formulas (\ref{eq8}) for the detonation wave
going from centre on the initial stage we have temperature about 1
GeV. The more strict analysis of the structure of scalaron
configurations presented below shows that the mass density of the
scalar field is proportional to $r^{-2}$. At the indicated central
density $\rho _{0}$\ the yield of the detonation wave on boundary
of configuration will give effective temperature behind the wave
front about 1 MeV. Further the dispersion of such inhomogeneous
macroscopic fireball is realized. This problem requires
independent research as well as the  problem of a relativistic
detonation with energy density decreasing on the indicated law.
But the rough estimate of average temperature of the whole
fireball with the indicated energy $E=Mc^{2}$ shows that it
achieves temperatures about 1 MeV at the size about 0.1 light
second. Then the electron - positron pairs which ''locked'' by
radiation disappear and the free dispersion of photons begins. The
reorganization of distribution of the density during the
dispersion can give an increasing of Doppler frequency of
radiation of photons in the head part of the fireball and leading
up it up to temperature about GeV again.

Now set of equations in the dimensionless form will be:
\begin{equation}
\label{eq16}
\frac{1}{\xi }\frac{d^{2}}{d\xi ^{2}}(\xi U)=-x^{2};\,\,\,\,\,\frac{1}{\xi }%
\frac{d^{2}}{d\xi ^{2}}(\xi x)+n^{2}Ux=0.
\end{equation}
Then $n$\ will be proportional to number of half-waves of the
field $\varphi $\ on the typical size of configuration. The
numerical solution of Eqns. (\ref{eq16}) is presented in Fig. 2.
\begin{figure}
\begin{picture}(500,290)
\put(-50,-520){\includegraphics{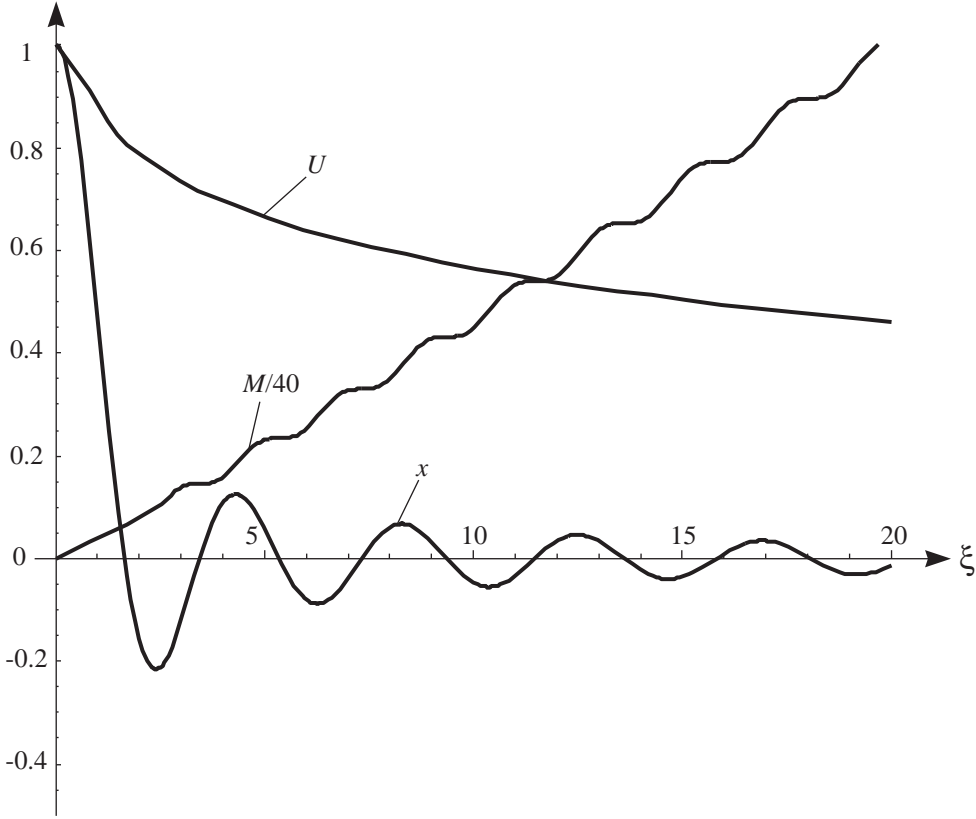}}
\end{picture}
\caption{\it The diagrams of dependence of dimensionless
gravitational potential $U$, potential of the scalar field $x$\
and total mass of configuration $M$\ on dimensionless radius $\xi
$.}
\end{figure}

In the equation on $x$\ there is the big parameter $n$\ that
allows to search for solution on $x$ as ~\cite{ref:Mig}:
\begin{equation}
\label{eq17}
x=f(\xi )\sin \Omega ;\,\,\,\,\,\Omega =n\int\limits_{0}^{\xi }\sqrt{U}%
d\zeta .
\end{equation}
The argument at a sine is the fast varying function in comparison
with the potential $U$. Then the solution of the second equation
from (\ref{eq16}) is possible to present as
\begin{equation}
\label{eq18}
x=\frac{1}{\xi U^{1/4}}\sin \Omega +\left(
\frac{1}{n^{2}}\right) .
\end{equation}
The obtained formula shows that according to expression
(\ref{eq13}) it corresponds to $\rho \sim r^{-2}$\ as was used in
estimations mentioned above. The substitution of Eqn. (\ref{eq18})
in the equation for potential gives
\begin{equation}
\label{eq19}
\frac{1}{\xi }\frac{d^{2}}{d\xi ^{2}}(\xi U)=-\frac{1}{2\xi ^{2}\sqrt{U}}%
\left[ 1-\cos 2\Omega \right] .
\end{equation}
Hence the gravitational potential consists of slowly varying part
described by the equation
\begin{equation}
\label{eq20}
\frac{1}{\xi }\frac{d^{2}}{d\xi ^{2}}(\xi
U)=-\frac{1}{2\xi ^{2}\sqrt{U}}
\end{equation}
and fast varying one $\stackrel{\sim}{U}$
\begin{equation}
\label{eq21}
\frac{d^{2}}{d\xi ^{2}}\left( \xi ^{2}\frac{d\stackrel{\sim}{U}}{%
d\xi }\right) =\frac{1}{2\sqrt{U}}\cos 2\Omega ,
\end{equation}
which in $1/n^{2}$\ times less then $U$\ as follows from
Eqn.~(\ref{eq21}).
For this reason it is not taken into account in the whole formulas with $%
\sqrt{U}$. Note that expression on the {\it r.h.s.} in Eqn.
(\ref{eq20}) is valid at $\xi \gg \xi_{0}\approx 1/n$.

\section{Conclusion}

\noindent The authors realize that the considered hypothesis
requires an detailed examination in many items. The mechanism of
creation of particles behind the front of the detonation wave is
not clear in details, though in a series of models there are
indication on a strong increasing of a mass of a field quantum at
an approaching to critical temperature~\cite{ref:Linde} that
inevitably should result in strong increasing of an oscillation
amplitude and creation of particles. The question with values of a
mass of quantums of scalar fields existing in the modern Universe
which generate dense cluster systems (''stars'') is not
clear~\cite{ref:Miel}. The desire to obtain the time of existence
of the detonation wave commensurable with observable duration of
bursts of GRB has forced to consider lengthy Newtonian scalaron
configurations. The problem of their existence and stability is
also requires original research. But if to refuse this hypothesis
it is possible to consider the ''detonation'' of relativistic
configurations. Then the finite fireball will be small and the
subsequent phenomenon of GRB will require considering its
expansion either in vacuum or in an exterior medium. Thus
according to available results there will be no problem with
insufficiency of an energy for an explanation of the whole
observable phenomenon. Anyway, if in the modern Universe there is
such unique form of a matter as fields of high density it would be
strange for Nature not to take advantage of a possibility to
convert their energy to radiation by an explosive way.

\par\bigskip
{\bf Acknowledgements}
\par\bigskip

\noindent Authors thank A. Pushkin and V. Belinskii for discussing
of this hypothesis on the initial stage of research.

\end{document}